\documentclass[pra,showpacs]{revtex4}
\usepackage{amsmath}
\usepackage[dvips,xdvi]{graphicx}
\usepackage{amssymb}
\usepackage{epsfig}
\newcommand{\be}{\begin{equation}}
\newcommand{\ee}{\end{equation}}
\newcommand{\bea}{\begin{eqnarray}}
\newcommand{\eea}{\end{eqnarray}}

\begin{document}

\title{Tkachenko modes and structural phase transitions of the vortex lattice of a two-component Bose-Einstein condensate}
\author{M. Ke\c{c}eli}
\author{M.~\"O.~Oktel}
\email{oktel@fen.bilkent.edu.tr}
\affiliation{ Department of Physics, Bilkent University, 06800 Ankara, Turkey }%

\date{\today}

\begin{abstract}
We consider a rapidly rotating two-component Bose-Einstein condensate (BEC) containing a vortex
lattice. We calculate the dispersion relation for small oscillations of vortex positions (Tkachenko
modes) in the mean-field quantum Hall regime, taking into account the coupling of these modes with
density excitations. Using an analytic form for the density of the vortex lattice, we numerically
calculate the elastic constants for different lattice geometries. We also apply this method to
calculate the elastic constant for the single-component triangular lattice.  For a two-component
BEC, there are two kinds of Tkachenko modes, which we call acoustic and optical in analogy with
phonons. For all lattice types, acoustic Tkachenko mode frequencies have quadratic wave-number
dependence at long-wavelengths, while the optical Tkachenko modes have linear dependence. For
triangular lattices the dispersion of the Tkachenko modes are isotropic, while for other lattice
types the dispersion relations show directional dependence consistent with the symmetry of the
lattice. Depending on the intercomponent interaction there are five distinct lattice types, and
four structural phase transitions between them. Two of these transitions are second-order and are
accompanied by the softening of an acoustic Tkachenko mode. The remaining two transitions are
first-order and while one of them is accompanied by the softening of an optical mode, the other
does not have any dramatic effect on the Tkachenko spectrum. We also find an instability of the
vortex lattice when the intercomponent repulsion becomes stronger than the repulsion within
components.

\end{abstract}

\pacs{03.75.Lm,03.75.Mn,03.75.Kk,67.40.Db}

\maketitle

\section{Introduction}
\label{intro}

One of the defining properties of superfluidity is that a superfluid responds to rotation by
forming quantized vortices. Generally, instead of forming multiply quantized vortices, it is more
favorable for a superfluid to create many singly quantized vortices and arrange them in a vortex
lattice. Since the original predication of such structures by Abrikosov \cite{abr57}, vortex
lattices have been observed in type-II superconductors \cite{tes67}, superfluid helium
\cite{ypa82}, Bose-Einstein condensed gases (BECs) \cite{hce01, arv01} and most recently in
ultracold fermion superfluids\cite{zas05}.

Once a vortex lattice is formed in a superfluid, small deviations of the vortices from their
equilibrium positions require relatively small energy compared to other hydrodynamic modes of the
system, and collective behavior of such small deviations result in a low-energy branch in the
excitation spectrum. The modes on this branch, which were studied by Tkachenko in the context of
superfluid helium \cite{tka66}, are called Tkachenko modes and in a simplified picture can be
thought of as phonons of the vortex lattice. Tkachenko modes strongly affect the dynamics of the
superfluid \cite{bch83}, and play an important role in many different problems, ranging from vortex
melting \cite{shm02} to neutron star glitches \cite{rud70}.

Recent experiments on ultracold atoms have been successful in creating large vortex lattices in
rotating harmonically trapped BECs \cite{hce01, arv01}. Remarkable results about vortex dynamics
have been obtained, including the observation of Tkachenko modes over a large range of rotation
frequencies \cite{ces03}. In this experiment, after the formation of the vortex lattice, a resonant
laser beam was focused on the center of the condensate to excite the Tkachenko modes and
subsequently their frequency was measured. As the rotation frequency is increased, a clear
reduction in the Tkachenko mode frequencies is observed.

Theoretical study of Tkachenko modes of trapped BECs has been carried out by a number of groups
\cite{acr02, bay03, son04, cst05, son05, cps04, bwc04, mkm04}. In particular, the effects of finite
size of the vortex lattice and the compressibility of the BEC lead to major differences in the
Tkachenko spectrum compared with the Tkachenko modes of an incompressible superfluid such as
helium. As the rotation frequency of the cloud is increased, the compressibility of the BEC starts
to play an important role, reducing the shear modulus of the vortex lattice and thus the Tkachenko
mode frequencies. When the rotation frequency $\Omega$ becomes close to the chemical potential $\mu
= g n$, the gas enters the mean-field quantum Hall regime \cite{ho01} where only the states in the
lowest Landau level (LLL) are populated. Here, the trend of decrease in Tkachenko frequencies
continues. As the rotation frequency $\Omega$ gets closer to the trapping frequency $\omega$, more
vortices enter the system, and mean-field description breaks down at the point where the number of
vortices is comparable to the number of particles \cite{shm02}. In the strongly correlated regime,
the vortex lattice is expected to melt into a vortex liquid and subsequently go through a sequence
of quantum Hall states ending with the Bosonic Laughlin state when $(\omega-\Omega)/\omega \sim
1/N$, with $N$ being the number of particles \cite{cwg01}.

In the experiments of the JILA group, rotation frequencies up to 99\% of the trapping frequency
have been achieved \cite{sce04} and a calculation of the Tkachenko frequencies in the mean-field
quantum Hall regime \cite{bay03} found good agreement with the observed frequencies. However, a
number of papers have since argued that this calculation uses an incorrect value for the shear
modulus of the vortex lattice \cite{son04,cst05}. When the recalculated value of the shear modulus,
which is an order of magnitude higher, is used, the experimental results seem to indicate that the
gas is not in the LLL regime. Although in this paper we mainly consider two-component BECs, our
method is applicable to the single-component lattice, and our calculations are in excellent
agreement with the latter value for the shear modulus, suggested by Sonin \cite{son04}.

The versatility of the trapped cold atom experiments have enabled the creation of new superfluids,
such as mixtures and spinor condensates. In a remarkable experiment the JILA group has been able to
create a two-component BEC and study its behavior under rotation \cite{sce04b}. The equilibrium
vortex lattice structures have been calculated by Mueller and Ho \cite{mho02}, and separately by
Kasamatsu, Tsubota, and Ueda \cite{ktu04}. Experimentally, an interlaced square lattice of
two-components has been observed. Furthermore, using an excitation procedure similar to the
one-component case, vortex lattice oscillations have been induced in the two-component BEC,
however, these excitations were found to be heavily damped and have not yielded a measurement for
Tkachenko frequencies. Motivated by this experiment, in this paper, we calculate the Tkachenko
modes of a two-component vortex lattice, and investigate the structural phase transitions between
different lattice geometries.

We consider a large two-component vortex lattice in the LLL regime. To simplify the calculations,
we assume that both components have the same density and same scattering length within each
component. As the scattering length between atoms from different components is varied, the vortex
lattice goes through structural phase transitions, forming five different lattice geometries
\cite{mho02}. For all these lattice geometries, we calculate the elastic constants of the vortex
lattice, and subsequently the dispersion relations for long-wavelength Tkachenko modes. Our main
results are summarized below.

Unlike a single-component vortex lattice, where there is only one branch of Tkachenko modes, the
two-component lattice has two branches. The situation is similar to phonon modes of a diatomic
solid compared with a monoatomic solid. When the number of atoms per unit cell is doubled, so are
the number of phonon modes. In analogy with phonons, we call these branches acoustic Tkachenko
modes, and optical Tkachenko modes. However, these names are not intended to imply that one branch
couples more strongly to light than sound, or vice versa. As a simple picture, one may think that
when an acoustic mode is excited two vortices inside the unit cell of the lattice oscillate
in-phase. In other words, acoustic modes are oscillations of the ``center of mass" of the unit
cell, while the vortices positions with respect to the center of mass remain stationary. For an
optical Tkachenko mode, vortices of different components oscillate in opposite phase, leaving the
``center of mass'' of each unit cell stationary. In this paper, we choose our interactions such
that there is symmetry under the exchange of components, which makes the above definitions of
optical and acoustic unambiguous. If this symmetry is broken, as is the case with the parameters of
the JILA experiment, there will still be two modes, but both of them will contain a mixture of
acoustic and optical behavior.

For an incompressible superfluid such as helium, or at low  rotation frequencies for BEC, Tkachenko
modes in a single-component vortex lattice have linear wave-vector dependence $\omega_T \propto k$
\cite{bch83}. However, when compressibility of the fluid becomes important, such as a BEC in the
LLL, Tkachenko modes are  quadratic  in the wave-vector $\omega_T \propto k^2$ \cite{bay03}. We
find that a similar softening happens for the two-component vortex lattice. For an incompressible
fluid, acoustic Tkachenko modes have linear long-wavelength behavior, while the optical Tkachenko
modes are gapped. For a two-component BEC in the LLL, acoustic Tkachenko modes have quadratic
wave-vector dependence $ \omega_{T}^{\rm{ac}} \propto k^2 $, while the optical modes are not gapped
any more, but have linear wave-vector dependence $\omega_{T}^{\rm{op}} \propto k$.

Another important property of the Tkachenko modes of a single-component system is their isotropy.
Tkachenko mode frequencies are independent of the  direction of the excitation wave vector
$\vec{k}$. This can be traced back to the fact that the underlying vortex lattice is triangular,
and similar to acoustic waves in a triangular lattice, Tkachenko modes have isotropic behavior
\cite{lli70}. For two-component vortex lattices, this behavior is not expected any more, and indeed
we find that when the underlying lattice has less than sixfold symmetry, both acoustic and optical
Tkachenko modes are anisotropic. In all cases the anisotropy reflects the reduced symmetry of the
lattice, giving fourfold symmetric dispersion relations for the square lattice, and twofold
symmetric spectra for rhombic and rectangular lattices.

Another interesting point about the two-component vortex lattices is the possibility of structural
phase transitions between different lattice geometries. For a two-component BEC in the LLL there
are five lattice structures and four structural phase transitions between them. Two of these are
continuous, second-order transitions, while the other two are first-order transitions. In
structural phase transitions of solids, second-order phase transitions are signalled by the
softening of an acoustic-phonon mode, while first-order transitions are usually, but not always,
accompanied by the softening of an optical-phonon mode. (A soft mode can be described as a branch
of excitation that has zero frequency over a large range of wave vectors \cite{kit96}.) We find
that a similar scenario plays out for the vortex lattices of two-component BECs, both second-order
phase transitions have a soft acoustic Tkachenko mode. Of the two first-order phase transitions,
one is accompanied by a soft optical Tkachenko mode, while the other does not have a direct effect
on the long-wavelength Tkachenko spectrum of the system.

There are two other instabilities in the two-component BEC system. When the intercomponent
attraction is stronger than the repulsion within each component, the gas is unstable towards
collapse. In the opposite limit, when the intercomponent interaction is repulsive and stronger than
the intracomponent repulsion we find an instability in the optical Tkachenko mode spectrum, most
possibly signaling a transition to a phase separated state.

The paper is organized as follows. In the next section, we introduce the Hamiltonian for the
two-component rotating gas in the LLL, and introduce the different lattice types that are found by
energy minimization. In Sec. \ref{elastic}, we outline our method of calculation for elastic
coefficients, and calculate the shear modulus of a one-component condensate as an example. In Sec.
\ref{equations}, we write the coupled equations for the vortex modes and density modes, which are
valid for all lattice types. In the next five sections \ref{ovrtriangle} -\ref{rhombus}, we
calculate the elastic energy for each lattice type, and by solving the coupled equations, we find
the dispersion relations of acoustic and optical Tkachenko modes. We also study the directional
dependence of the dispersion relations and the polarization of the Tkachenko modes for each lattice
type. In Sec. \ref{phasetransitions}, we discuss  the structural phase transitions and identify the
soft modes associated with each transition.  Finally, in Sec. \ref{conclusion}, we summarize our
results and discuss their consequences for experiments.

\section{Vortex lattices of two-component BEC}
\label{lattices}

In this section, we consider the equilibrium vortex lattice configurations of a two-component BEC.
This problem has been studied in the LLL regime analytically by Mueller and Ho \cite{mho02}, and
for general rotation frequencies numerically by Kasamatsu {\it et al.} \cite{ktu04}. We confine
ourselves to the LLL and our method of calculation of the elastic constants relies on the analytic
approach developed by Mueller and Ho.

We consider a two-component BEC in a quadratic trap with trapping frequency $\omega$. The trap
frequency, the mass of the particles $m$, and the total number of particles are assumed to be the
same for both components. We take the gas to be rotating at frequency $\Omega$, and assume that the
total number of particles in each component is large enough to form a large vortex lattice without
a breakdown of the mean-field description of the system. Furthermore, we assume that the scattering
lengths of the particles are such that, interaction parameters satisfy
\bea
g_{11}&=&g_{22}=g, \\ \nonumber g_{12}&=& \alpha g.
\eea
We investigate the behavior of vortex lattice geometry and the Tkachenko modes as the ratio of
intercomponent scattering length to intracomponent scattering length $\alpha$ is varied,
\be
\alpha=\frac{g_{12}}{g}.
\ee

We limit our discussion of vortex lattices to two dimensions, assuming that the vortex lattice is
not modified along the rotation axis. This assumption is not very restrictive, as it has been shown
that if the cloud is sufficiently broad in the third dimension, vortex bending is negligible except
at the edges of the cloud \cite{ho01}. In the opposite limit of a two-dimensional condensate, our
approach is formally valid, however, mean-field theory may not be reliable for such a system. The
energy functional for our system can be written as
\be
E= \sum_{i=1,2} \int d^2r \Psi_i^*(r) \left( - \frac{\hbar^2}{2m} \nabla^2 + \frac{1}{2} m \omega^2
r^2 - \Omega L_z \right) \Psi_i(r) + V_{\rm{int}}.
\ee
Here $\Psi_i$ is the wave function of component $i$, $L_z$ is the angular momentum along the
rotation direction and $V_{\rm{int}}$ is the interaction energy given as
\be
V_{\rm{int}}= g \int d^2r \left(\frac{1}{2} \left[ |\Psi_1(r)|^4 + |\Psi_2(r)|^4 \right] + \alpha
|\Psi_1(r)|^2 |\Psi_2(r)|^2 \right).
\ee

When the rotation frequency is close enough to the trapping frequency the particles can only
populate levels in the LLL. For such a gas, which is in the mean-field quantum Hall regime, the
wave functions have the form
\be
\Psi_i(r) = f_i(z) e^{-\frac{z \bar{z}}{2 \sigma^2}},
\ee
where $z=x+iy$ and $\sigma$ is the radius of the cloud. The requirement of analyticity on the wave
function essentially determines the form of the wave function in terms of the positions of the
vortices (up to an entire function with no zeros). Thus it is possible to introduce a variational
wave function, using just the lattice basis vectors as variational parameters.

For a two-component BEC, when both of the components are rotating at the same frequency, vortex
lattices in each component have the same lattice structure, but are shifted from each other. Thus
in the LLL, we can determine the wave functions for both components in terms of just three,
two-dimensional vectors $\vec{a}_1$ and $\vec{a}_2$, the basis vectors of the lattice, and
$\vec{d}$, the offset between the two lattices. Thus the vortices of the first component are at
$\vec{r}_{1,n,m}=n \vec{a}_1 + m \vec{a}_2$, with $n,m$ integers, while the vortices of the second
component are at $\vec{r}_{2,n,m}=n \vec{a}_1 + m \vec{a}_2 + \vec{d}$. Although we need six real
numbers to describe these three vectors, the actual number of variational parameters is lower,
namely 4. First of all, the vortex density $\nu_c^{-1}$ is fixed by rotation frequency $\Omega$,
thus it is possible to fix the length of one of the vectors and scale all others by this length.
Second, the rotational symmetry of the problem permits one to fix the overall orientation of the
vectors. We choose the first lattice basis vector $\vec{a}_1$ to lie along the $\hat{x}$ direction,
and denote its length as $a_1$. The remaining two vectors can then be written as
\bea
\vec{a}_2&=& a_1(u \hat{x}+v \hat{y}), \\ \nonumber \vec{d}&=& a_1[(a+b u) \hat{x}+ b v \hat{y}].
\eea
The variational calculation is made in terms of the dimensionless parameters $u, v, a, b$, and then
the length $a_1$ is fixed by requiring the wave function to have correct density of vortices $\nu_c
= a_1^2 v$.

Once the positions of the vortices are known one can write the variational wave function as a
Jacobi Elliptic function $\Theta$, or as one of the doubly periodic functions that are related to
the $\Theta$ up to an entire function, such as the $\sigma$ function or the modified $\zeta$
function \cite{tka66}. In terms of the Jacobi theta function we can write
\bea \Psi_1(z)&=& N_1 \Theta(\zeta,\tau) \exp(\frac{\pi z^2}{2
\nu_c}-\frac{z \bar{z}}{2 l}), \\ \nonumber \Psi_2(z)&=& N_2 \Theta(\zeta-(a+b u+i b v),\tau)
\exp(\frac{\pi z^2}{2 \nu_c}-\frac{z \bar{z}}{2 l}).
\eea
Here $\zeta=z/a_1$, $\tau=u+i v$, $l=\sqrt{\hbar/m \omega}$, and $N_1,N_2$ are normalization
constants to be determined. With these wave functions we calculate the densities of the
two-components as
\bea
|\Psi_1(\vec{r})|^2 &=& C g(\vec{r}) e^{-r^2/\sigma^2}, \\ \nonumber |\Psi_2(\vec{r})|^2 &=& C
g(\vec{r}-\vec{d}) e^{-r^2/\sigma^2},
\eea
where the function $g$ is periodic with lattice vectors,
\be
g(\vec{r}+n\vec{a_1} + m \vec{a_2})= g( \vec{r}),
\ee
for all integers $n,m$. The periodic part of the density admits a Fourier series representation in
terms of the reciprocal-lattice basis vectors,
\be
g(\vec{r})= \frac{1}{\nu_c} \sum_{\vec{K}} g_{\vec{K}} e^{i \vec{r}\cdot\vec{K}},
\ee
with the sum carried out over all reciprocal lattice points generated by $\vec{K}_1= \frac{2
\pi}{\nu_c} \vec{a}_2 \times \hat{z}$ and $\vec{K}_2= -\frac{2 \pi}{\nu_c} \vec{a}_1 \times
\hat{z}$. The utility of using the Jacobi theta function is that the Fourier components of
$g_{\vec{K}}$ can be calculated with relative ease. For $\vec{K}=m_1 \vec{K_1} + m_2 \vec{K_2}$ one
has
\be
g_{\vec{K}}= (-1)^{m_1+m_2+m_1 m_2} e^{-\frac{\nu_c \vec{K}^2}{8 \pi}} \sqrt{\frac{\nu_c}{2}}.
\ee

In the LLL, the lattice structure is entirely determined by the interaction energy. For the
parameters used here minimization of the interaction energy reduces to a minimization of the
following simple quantity with respect to $u,v,a$, and $b$:
\be
J= \sum_{\vec{K}} \left| \frac{g_{\vec{K}}}{g_{\vec{0}}} \right|^2 \left(1+ \alpha \cos(\vec{K}
\cdot \vec{d})\right).
\ee
It must be noted that this expression is obtained in the limit of a very large vortex lattice,
formally setting the cloud radius $\sigma$ to infinity. The minimization of $J$ is done numerically
with considerable ease as the Fourier coefficients of the density $g_{\vec{K}}$ are known
analytically. For each value of $\alpha=g_{12}/g$, $J$ can be calculated by truncating the rapidly
converging sum to the desired accuracy, and the values $u_*, v_*, a_*, b_*$ that minimize $J$ can
be found. These values determine the lattice geometry for each component and also the offset of the
lattices of two-components.

As the ratio of the intercomponent interaction to intracomponent interaction $\alpha$ is varied,
five different lattice types are found to minimize the interaction energy. Here, we give a brief
description of each lattice, and in Secs. \ref{ovrtriangle} - \ref{rhombus}, the Tkachenko spectrum
for each lattice type is calculated.

\begin{figure}
\centering\includegraphics[scale=0.3]{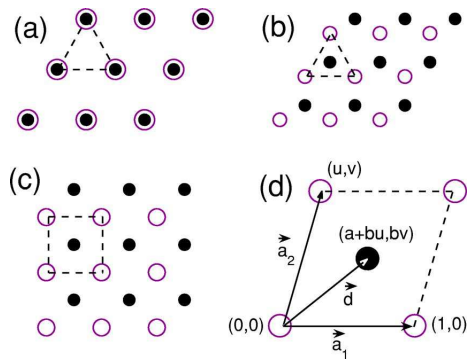} \caption[]{Lattice geometry for an overlapped
triangular lattice (a), an interlaced triangular lattice (b), and a square lattice (c). Unit cells
are shown with dashed lines. (d) Unit cell geometry for an arbitrary lattice. White and black dots
represent vortices of different components. Definitions of $a, b, u, v$ are given in Sec.
\ref{lattices}.} \label{latticegeo}
\end{figure}

When the interaction between the two-components is attractive, i.e., $\alpha<0$, the system
minimizes its energy by positioning the vortex lattices of two-components on top of each other.
However, for very large attraction, $\alpha<-1$, there is an instability towards collapse. In the
range $-1<\alpha<0$, both components form triangular lattices which overlap with each other (See
fig. \ref{latticegeo}). This overlapped triangular lattice is described by the parameters $u_*=1/2,
v_*=\sqrt{3}/2, a_*=b_*=0$, which do not change with $\alpha$ in the given range. We find, however,
that the elastic constants of the lattice depend on $\alpha$, and so do the Tkachenko modes.


\begin{figure}
\centering\includegraphics[scale=0.25]{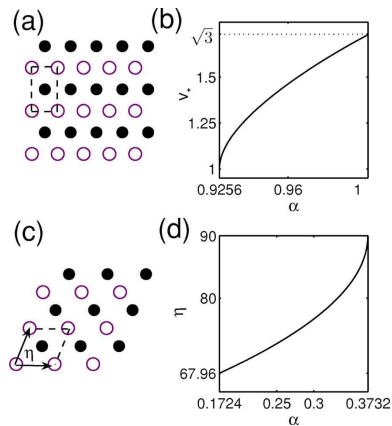} \caption[]{(a) Lattice geometry for a rectangular
lattice; a unit cell is shown with dashed lines. (b) Change of the aspect ratio of the rectangle
$v_*$ with respect to interaction strength $\alpha$. The unit cell grows in the $y$ direction as
$\alpha \to 1 $. At $\alpha=1$, $ v_*=\sqrt{3}$. (c) Lattice geometry of a rhombic lattice, dashed
lines showing a unit cell. $\eta$ is twice the opening angle of the rhombic unit cell. (d) Plot of
$\eta$ vs $\alpha$ for the rhombic lattice. As $\alpha \to 0.3732 $, $\eta \to 90^\circ$, and the
rhombus continuously changes to a square. At $\alpha=0.1724 $, $ \eta$ makes a jump from $60^\circ$
to $67.958^\circ$.} \label{latticegeo2}
\end{figure}

If the intercomponent interaction becomes repulsive, it is no more favorable to put the two vortex
lattices on top of each other. Instead, the most favorable places to put the vortices of
one-component would be the density maxima of the other component. This simple insight holds true
for all lattice types found by the minimization procedure, however the lattice type of each
component changes as $\alpha$ is varied. For weak repulsion between the components,
$0<\alpha<0.1724$, each component forms a triangular lattice. Within a unit cell, there is more
than one density maximum, so it would seem that there are multiple positions for the vortex lattice
of the second component to be placed. However, these positions are related with the overall
symmetry of the lattice, so the minimization procedure gives the lattice parameters $u_*=1/2,
v_*=\sqrt{3}/2 , a_*=b_*=1/3$. Again, the overall structure of this interlaced triangular lattice
(see Fig. \ref{latticegeo}) does not change with $\alpha$. At $\alpha=0.1724$, there is a
first-order phase transition from an interlaced triangular lattice to a rhombic lattice. In the
range $0.1724<\alpha<0.3733$, the unit cell of vortex lattices of each component are rhombuses. The
vortex lattice of one-component is placed at the center of the rhombuses formed by the vortex
lattice of the other component. The angle of the rhombus $\eta$ (see Fig. \ref{latticegeo2}),
varies continuously from $67.9^\circ$ to $90^\circ$, while the offset remains the same,
$a_*=b_*=1/2$.  At $\alpha=0.3733$, there is a second-order phase transition to a square lattice.
In the range $0.3733<\alpha<0.9256$, the lattice is parametrized by $u_*=0, v_*=1, a_*=b_*=1/2$
(see Fig. \ref{latticegeo2}). As the interaction is increased further, there is a second-order
phase transition to a rectangular lattice at $\alpha=0.9256$. In a rectangular lattice, vortices of
one-component are always found at the centers of the rectangles formed by the vortices of the other
component, i.e., $a_*=b_*=1/2$. However, the aspect ratio of the rectangle increases continuously.


For a nonrotating system, there is a phase-separation instability at $\alpha=1$. This instability
is not found in the results of the energy minimization described above. However, when the coupling
between the density oscillations and vortex motion is taken into account, as in Sec.
\ref{rectangle}, it is found that at this point there is an instability. Thus the system is not
described by a  vortex lattice beyond $\alpha=1$.

After a survey of the possible lattice structures and the analytic method that is used to find
these structures, in the next section we describe how the same analytic approach can be used to
calculate the elastic constants of the discussed vortex lattices.

\section{Numerical calculation of elastic constants}
\label{elastic}
The power of the analytic approach introduced in the previous section is that it can also be used
to calculate the energies of lattice structures which are slightly deformed from the minimum-energy
configuration. As the numerical calculation of the energy for a given lattice is quite simple, it
is possible to evaluate the energy for configurations where lattice parameters have small
deviations from their minimum-energy values. Such small deviations can also be described by a
hydrodynamic approach. Assuming that the lattice deformations are sufficiently smooth, the vortex
lattice can be  treated as an elastic medium. The form of the elastic energy is constrained by
symmetries of the  lattice, and for small deformations, can always be taken as quadratic in
displacements. Thus the long-wavelength behavior of the lattice is described by an elastic energy
that is quadratic in the vortex displacement field, and the problem reduces to the calculation of
the elastic constants, which are the coefficients of the quadratic terms in vortex displacements.

Our approach is to numerically calculate the energy of the vortex lattice close to the equilibrium
position, and then find the elastic coefficients of the vortex lattice by making quadratic fits to
the calculated energy. As a demonstration of this method, we first calculate the shear modulus of
the triangular lattice of a one-component BEC. A vortex lattice in a one-component BEC is
parametrized by two two-dimensional vectors $\vec{a}_1,\vec{a}_2$, the lattice basis vectors. The
lattice basis vectors define the equilibrium positions of the vortices, and we denote the deviation
of the vortex at lattice site $n,m$, from its equilibrium by the vector $\vec{\epsilon}_{n,m}$. So
the position of the vortex $\vec{r}_{n,m}$ is
\be
\vec{r}_{n,m}= n \vec{a}_1 + m \vec{a}_2 + \vec{\epsilon}_{n,m}.
\ee
If the vortex displacements are sufficiently smooth over large length scales, one can describe a
long-wavelength vortex displacement field $\vec{\epsilon}(x,y)= \epsilon_x(x,y)
\hat{x}+\epsilon_y(x,y) \hat{y}$ by a suitable coarse-graining procedure. For a triangular lattice,
the elastic energy density can then be written as
\be
\label{epsilonelastic}
\varepsilon_{\rm{elastic}}= C_1 \left( \frac{\partial \epsilon_x}{\partial x} + \frac{\partial
\epsilon_y}{\partial y} \right)^2+ C_2 \left[ \left(\frac{\partial \epsilon_x}{\partial x} -
\frac{\partial \epsilon_y}{\partial y} \right)^2
 +\left(\frac{\partial \epsilon_x}{\partial y} + \frac{\partial \epsilon_y}{\partial x} \right)^2 \right].
\ee
For a gas in the LLL, compression modulus is zero, $C_1=0$ \cite{bch83}. Using the analytic method
introduced above, the shear modulus $C_2$ is determined as follows.

\begin{figure}
\centering\includegraphics[scale=0.4]{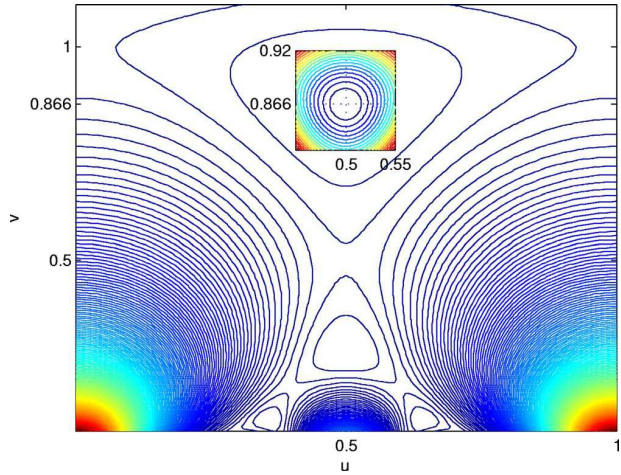}
\caption[]{ Contour plot of the energy for a one-component triangular lattice, Eq.
(\ref{energy}). The inset is a closer view around the equilibrium point. Circular contours indicate
that quadratic fit of Eq. (\ref{uvelastic}) is possible.}
\label{contour}
\end{figure}

For a one-component vortex lattice, the minimum-energy configuration is found by minimizing
\be\label{energy}
I= \sum_{\vec{K}}\left| \frac{g_{\vec{K}}}{g_{\vec{0}}} \right|^2,
\ee
and yields $u_*=1/2, v_*=\sqrt{3}/2$, the triangular lattice. We calculate the energy around this
point by varying $u$ and $v$ from their equilibrium values. A contour plot of the energy around the
equilibrium point is given in Fig. \ref{contour}. To this form we can successfully fit a quadratic
form, giving us an elastic energy of the form
\be
\label{uvelastic}
E_{\rm{elastic}}= \frac{g n^2}{2} \left[ C_u (u-u_*)^2 + C_v (v-v_*)^2 +C_{uv} (u-u_*)(v-v_*)
\right],
\ee
and determine
\bea
C_u&=&0.3177, \\ \nonumber C_v&=&0.3177, \\ \nonumber C_{uv}&=&0.000.
\eea

The fact that $C_u=C_v $ shows that there is only one shear modulus for a triangular lattice, and
validates our numerical procedure. To find the connection between $C_u$ and the shear modulus
$C_2$, we must determine the displacement field corresponding to small changes of the lattice basis
vectors. By taking into account that the unit-cell volume is fixed by rotational frequency, the
correspondence between $u,v$ and vortex displacement field is found as
\bea
\frac{\partial \epsilon_x}{\partial y} &=& \frac{u-u_*}{v_*}, \qquad \frac{\partial
\epsilon_x}{\partial x} = -\frac{v-v_*}{2 v_*}, \\ \nonumber \frac{\partial \epsilon_x}{\partial y}
&=& \frac{v-v_*}{2 v_*}, \qquad \frac{\partial \epsilon_x}{\partial y} = 0.
\eea

By substituting these expressions in Eq. (\ref{epsilonelastic}), using $u_*=1/2, v_*=\sqrt{3}/2$
and comparing with Eq. (\ref{uvelastic}) we obtain
\be
C_2 = \frac{3}{8} C_u = 0.1191 g n^2.
\ee
This value is an order of magnitude larger than the value used by Baym \cite{bay03}, and is in
excellent agreement with Sonin \cite{son04}. This is not surprising, as our method of obtaining the
shear modulus is equivalent to the deformation of the lattice used by Sonin.  However, the
simplicity of our numerical method enables us to calculate the elastic coefficients of more complex
lattices, such as the two-component lattices discussed in this paper.

For a two-component lattice, the energy of the lattice depends not only on the lattice basis
vectors, but also on  the offset of two lattices from each other. Hence there are four variational
parameters $u, v, a, b$. Elastic energy around the minimum-energy point has to be expressed as a
quadratic form in all of these variables. We numerically calculate the energy of the lattice for
many points around the minimum-energy point and then express the elastic energy by fitting to a
form
\be
\label{uvabelastic}
E_{\rm{elastic}}= \frac{g n^2}{2} \left[ C_u (u-u_*)^2 + C_v (v-v_*)^2 +C_{uv} (u-u_*)(v-v_*) + C_a
(a-a_*)^2+ C_b (b-b_*)^2 + C_{ab} (a-a_*)(b-b_*) \right].
\ee
Here, due to the symmetry between component 1 and component 2, it is not necessary to include terms
that mix displacements $a,b$ with deformations of the lattice $u,v$. This is essentially the
decoupling of optical Tkachenko modes from acoustic Tkachenko modes as discussed in Sec.
\ref{intro}.

In the next section, we give the hydrodynamic equations for a two-component condensate, but leave
the form of the elastic energy unspecified. In the following sections, the form of the elastic
energy and the values of the elastic constants are given separately for each lattice type. After
the elastic energy is specified, hydrodynamic equations are solved and the dispersion relations for
Tkachenko modes are obtained.

\section{Hydrodynamic equations}
\label{equations}

The oscillations of vortices about their equilibrium positions can be described by a hydrodynamic
theory by treating the vortex lattice as an elastic medium. For trapped BECs it is important to
take into account the compressibility of the gas, as the vortex lattice oscillations are coupled to
density oscillations in a nontrivial way. The superfluid hydrodynamics that takes this effect into
account has been developed by a number of groups in the context of superfluid He \cite{bch83}, and
more recently applied to rotating BECs by Baym \cite{bay03}. Here we describe the hydrodynamics of
a two-component vortex lattice by generalizing this hydrodynamics to a two-component BEC.

As the hydrodynamic variables, we use the densities of each component $n_i(\vec{r},t)$,
corresponding velocity fields $\vec{v}_i(\vec{r},t)$, and the vortex displacement fields
$\vec{\epsilon}_i(\vec{r},t)$ introduced in the previous section. Here $i=1,2$ is component index,
giving us a total of six hydrodynamic fields. However, not all of these fields are independent, as
is apparent in the calculation below. We also set $\hbar=1$ in the calculation for convenience.

The long-wavelength average of the velocity field is not irrotational, but is linked to the
compressions of the vortex lattice,
\be
\label{vorticity}
\vec{\nabla} \times \vec{v}_i = - 2 \Omega \vec{\nabla} \cdot \vec{\epsilon}_i, \qquad i=1,2.
\ee
Similarly the superfluid acceleration equation holds for each component
\be
\label{acceleration} m \left( \frac{\partial \vec{v}_i}{\partial t} + 2 \vec{\Omega} \times
\frac{\partial \vec{\epsilon}_i}{\partial t} \right) = - \vec{\nabla} \mu_i, \qquad i=1,2.
\ee
Here $\mu_i$ is the chemical potential of component $i$. Below we leave the index $i$ unspecified
to indicate that the equation is valid for both components.

The conservation of particle number results in the continuity equation
\be
\label{continuity}
\frac{\partial n_i}{\partial t} + \vec{\nabla} \cdot \left( n_i \vec{v}_i \right) = 0,
\ee
while momentum conservation gives
\be
\label{momentum} m \left( n_i \frac{\partial \vec{v}_i}{\partial t} + 2 n_i \vec{\Omega} \times
\vec{v}_i \right) + \vec{\nabla} P_i = -\vec{\sigma}_i.
\ee
Here $P_i$ is the pressure, related to the chemical potential as $\vec{\nabla} P_i = n_i
\vec{\nabla} \mu_i$, and for a weakly interacting two-component condensate satisfies
\bea
\vec{\nabla} P_1 &=& g n \vec{\nabla} n_1 + \alpha g n \vec{\nabla} n_2, \\ \nonumber \vec{\nabla}
P_2 &=& g n \vec{\nabla} n_2 + \alpha g n \vec{\nabla} n_1.
\eea
The stress vectors $\sigma_i$ are obtained by taking the functional derivative of the elastic
energy with respect to vortex displacement fields, as in elasticity theory,
\be
\label{stress}
\vec{\sigma}_i = \frac{\delta E_{\rm{elastic}}}{\delta \vec{\epsilon}_i}.
\ee

Using Eqs. (\ref{momentum}) and (\ref{vorticity}), we have
\be
2 m \vec{\Omega} \times \left( \frac{\partial \vec{\epsilon}_i}{\partial t} - \vec{v}_i \right) =
\frac{\vec{\sigma}_i}{n}.
\ee
The curl and divergence of these equations lead to
\be
\vec{\nabla} \cdot \left( \frac{ \partial \vec{\epsilon}_i}{\partial t} - \vec{v}_i \right) =
\frac{\vec{\nabla} \times \vec{\sigma}_i}{2 \Omega n m},
\ee
and
\be
\vec{\nabla} \times \frac{\partial \vec{\epsilon}}{\partial t} + 2 \Omega \vec{\nabla} \cdot
\vec{\epsilon}_i = - \frac{ \vec{\nabla} \cdot \vec{\sigma}_i}{2 \Omega n m}.
\ee
Similarly, the divergence of the superfluid acceleration equation gives
\bea
\left( -\frac{\partial^2}{\partial t^2} + \frac{g n}{m} \nabla^2 \right) n_1 + \alpha \frac{g n}{m}
\nabla^2 n_2 &=& 2 \Omega n \vec{\nabla} \times \frac{\partial \vec{\epsilon_1}}{\partial t}, \\
\nonumber \left( -\frac{\partial^2}{\partial t^2} + \frac{g n}{m} \nabla^2 \right) n_2 + \alpha
\frac{g n}{m} \nabla^2 n_1 &=& 2 \Omega n \vec{\nabla} \times \frac{\partial
\vec{\epsilon_2}}{\partial t}.
\eea

At this stage it  is preferable to take advantage of  the symmetry of the equations under the
exchange of component 1 with component 2. We define the symmetric and antisymmetric variables as
\bea
n_+ &=& n_1 + n_2, \qquad \vec{\epsilon}_+ = \vec{\epsilon}_1 + \vec{\epsilon}_2, \qquad
\vec{\sigma}_+ = \vec{\sigma}_1 + \vec{\sigma}_2,
\\ \nonumber
n_- &=& n_1 - n_2, \qquad \vec{\epsilon}_- = \vec{\epsilon}_1 - \vec{\epsilon}_2, \qquad
\vec{\sigma}_- = \vec{\sigma}_1 - \vec{\sigma}_2.
\eea

In terms of these variables we obtain two sets of three equations, where each set is decoupled from
the other. The polarization of the Tkachenko modes are controlled by the polarization equation
\be
\label{polarization}
\vec{\nabla} \times \frac{\partial \vec{\epsilon}_\pm}{\partial t} + 2 \Omega \vec{\nabla} \cdot
\vec{\epsilon}_\pm = - \frac{1}{2 m n \Omega} \vec{\nabla} \cdot \vec{\sigma}_\pm.
\ee
The usual sound mode equations for a two-component fluid are modified by the dynamics of the vortex
lattice as
\be
- \frac{\partial^2 n_\pm}{\partial t^2} + (1 \pm \alpha) \frac{g n }{m} \nabla^2 n_\pm = 2 n \Omega
\vec{\nabla} \times \frac{\partial \vec{\epsilon}_\pm}{\partial t}.
\label{sound}
\ee
The dynamics of the vortex lattice and its interaction with the density modes is governed by
\be
\label{lattice-sound}
\vec{\nabla} \cdot \frac{ \partial^2 \vec{\epsilon}_\pm}{\partial t^2} + \frac{1}{n}
\frac{\partial^2 n_\pm}{\partial t^2} = \frac{1}{2 n m \Omega} \frac{\partial}{\partial t}
\vec{\nabla} \times \vec{\sigma}_\pm.
\ee
Equations (\ref{polarization})-(\ref{lattice-sound}) form a linear set of six equations. However,
as the stresses $\sigma_\pm$, depend only on the lattice displacements $\epsilon_\pm$, with the
same sign, the three symmetric variable ($+$) equations are decoupled from the antisymmetric
variable ($-$) equations. Thus the ``$+$'' set describes the acoustic Tkachenko modes and their
coupling with the ``in-phase'' sound mode, while the ``$-$'' set describes the optical Tkachenko
modes and their coupling to the ``out-of-phase'' sound mode.

In the following sections, we specify the elastic energy $E_{\rm{elastic}}$ for each lattice type,
and calculate the dispersion relations of both acoustic and optical Tkachenko modes. Each section
starts with a brief description of the properties of the lattice type under consideration.
Subsequently we give the form of the elastic energy for this lattice type and the values of the
numerically calculated elastic constants. We then outline the solutions of the Tkachenko mode
equations Eqs. (\ref{polarization})-(\ref{lattice-sound}), for the specific form of the elastic
energy, and derive the dispersion relations of the acoustic and optical modes. Each section is
concluded by a discussion of the properties of the dispersion relation.

\section{Overlapped triangular lattice}
\label{ovrtriangle}

When the interaction between the two-components is attractive, it is energetically preferable to
have the density minima of the two-components to coincide. However, if the intercomponent
attraction is too strong there will be a collapse type instability. This insight is validated by
the calculations mentioned in Sec. \ref{lattices}, where for $-1< \alpha < 0$, the equilibrium
lattice structure is triangular for both components and the vortex lattices of the two-components
coincide. This overlapped triangular lattice is described by
\be
u_*=1/2,\qquad v_*=\sqrt{3}/2,\qquad a_*=b_*=0.
\ee

The elastic energy in all the vortex lattices can be separated into two parts, elastic energy due
to acoustic displacements $\epsilon_+$, and elastic energy due to optical displacements
$\epsilon_-$. There will not be any terms that contain both, as such contributions to energy change
sign under the exchange of components. So we can write
\be
E_{\rm{elastic}}= E_{\rm{elastic}}^{\rm{ac}} + E_{\rm{elastic}}^{\rm{op}}.
\ee
The acoustic contribution to the elastic energy will have the same form that is valid for a
triangular lattice. In the LLL the hydrostatic compression modulus is zero and we need to consider
only the shear modulus,
\be
E_{\rm{elastic}}^{\rm{ac}} = \int d^2r C^{\rm{ac}} \left[ \left( \frac{\partial
\epsilon_{+}^x}{\partial x} - \frac{\partial \epsilon_{+}^y}{\partial y}
 \right)^2 +\left( \frac{\partial \epsilon_{+}^x}{\partial y} + \frac{\partial \epsilon_{+}^y}{\partial x} \right)^2 \right].
\ee
Similarly, the only quadratic form one can make from $\epsilon_-$ which does not break the sixfold
symmetry of the lattice is
\be
E_{\rm{elastic}}^{\rm{op}}= \int d^2r C^{\rm{op}} \left(\vec{\epsilon}_{-}\right)^2.
\ee
The two elastic constants, $C^{\rm{ac}}$ and $C^{\rm{op}}$, control the acoustic and optical
Tkachenko modes, respectively. These two constants, however, have different dimensions, as is clear
from their definition. We first nondimensionalize these constants as follows.

For the acoustic shear modulus, we can define a dimensionless quantity $\tilde{C}^{\rm{ac}}$,
\be
\tilde{C}^{\rm{ac}}= \frac{C^{\rm{ac}}}{g n^2}.
\ee
As explained in Sec. \ref{elastic}, we can fit the energy near the minimum to a quadratic form,
\be
E_{\rm{elastic}}^{\rm{ac}}= \frac{1}{2} g n^2 C_u \left[ (u-u^*)^2+ (v-v_*)^2 \right],
\ee
which yields for the triangular lattice with $v_*=\sqrt{3}/2$
\be
\tilde{C}^{\rm{ac}} = \frac{3}{8} C_u.
\ee

\begin{figure}
 \centering\includegraphics[scale=0.35]{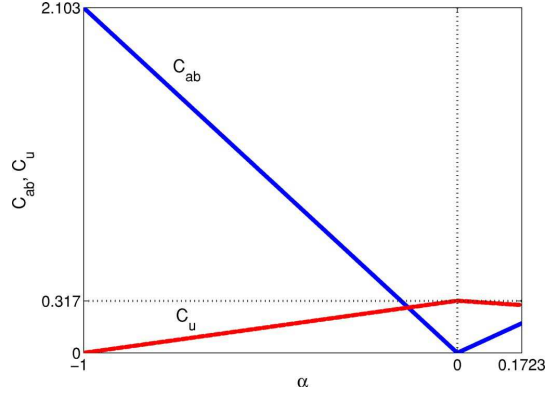}\caption[]{Elastic constants ($C_{ab}, C_u$) of overlapped ($-1<\alpha<0$) and interlaced ($0<\alpha<0.1723$) triangular lattices
with respect to $\alpha$. As the attraction between the components increases ($\alpha\to-1$), $
C_{ab}$ increases, and $C_u$ decreases linearly. When there is no interaction between components
($\alpha=0$), $ C_{ab}=0$ which causes the discontinuity in the transition to interlaced triangular
lattice. At $\alpha=0$, the value of $C_u$ is equal to the shear modulus of a one-component vortex
lattice.} \label{elastic1}
\end{figure}
The results of numerical calculation for $C_u$ are displayed in Fig. \ref{elastic1}. At $\alpha=0$,
the shear modulus for the acoustic modes takes the single-component value (per component) as
expected from two noninteracting vortex lattices. As $\alpha$ is decreased towards $-1$, the shear
modulus decreases linearly, signalling the collapse instability expected due to attractive
interaction between the components.

Similarly the optical elastic constant can be nondimensionalized as
\be
\tilde{C}^{\rm{op}}= \frac{d^2}{g n^2} C^{\rm{op}},
\ee
where $d$ is the lattice constant for the triangular lattice, which is related to rotation
frequency as
\be
d^2= \frac{2 \pi}{\sqrt{3} m \Omega}.
\ee
The optical part of the elastic energy can be fitted to the rotationally invariant form
\be
E_{\rm{elastic}}^{\rm{op}}=\frac{g n^2}{2} C_{ab} { [ (a-a_*)+ u_* (b-b_*) ]^2 + [ v_* (b-b_*) ]^2
}.
\ee
which results in
\be
\tilde{C}^{\rm{op}}=\frac{1}{2} C_{ab}.
\ee
The result of numerical calculation for $C_{ab}$ is plotted in Fig. \ref{elastic1}. As $\alpha$ is
decreased towards $-1$, it gets harder to separate the vortices of two-components, as expected from
the increasing attraction between the components.

Once the elastic constants are known, the calculation of the Tkachenko modes for different lattices
are straightforward, albeit tedious. In this section, we give a detailed calculation, while for all
other lattice types we simply present the results of the calculation.

We first start with the calculation of the acoustic Tkachenko modes. With the form of the elastic
energy given above, the acoustic stress is
\be
\vec{\sigma}_+ = - 4 C^{\rm{ac}} \nabla^2 \vec{\epsilon}_+.
\ee
Fourier transforming, we get $\vec{\sigma}_+ = 4 C^{\rm{ac}} k^2 \vec{\epsilon}_+$. Now, we also
Fourier transform the polarization equation (\ref{polarization}) to obtain
\be
\epsilon^y_+ = \frac{N_1}{D_1} \epsilon^x_+,
\ee
with
\bea
N_1 &=& -i \omega k_y - 2 \Omega k_x - \frac{ 2 C^{\rm{ac}}}{m n \Omega} k^2 k_x, \\ \nonumber D_1
&=& -i \omega k_x + 2 \Omega k_y + \frac{ 2 C^{\rm{ac}}}{m n \Omega} k^2 k_y.
\eea
Substituting the above result into Fourier transforms of Eqs. (\ref{sound}) and
(\ref{lattice-sound}), we obtain
\be
\left(\omega^2 - (1+\alpha) \frac{g n}{m} k^2 \right) n_+ + n \left( 4 \Omega^2 + \frac{4
C^{\rm{ac}}}{nm} k^2 \right) \frac{\omega k^2}{D_1} \epsilon_+^x =0
\ee
and
\be
- \frac{\omega^2}{n} n_+ + \left( \frac{ 4 C^{\rm{ac}}}{nm} k^2 - \omega^2 \right) \frac{ \omega
k^2}{D_1} \epsilon_+^x = 0.
\ee
By scaling the wave vector in frequency units,
\be
k' = \sqrt{\frac{g n}{m}} k,
\ee
and using dimensionless elastic constants we obtain the characteristic equation as
\be
\omega^4 - \omega^2 \left[ 4 \Omega^2 + (1+\alpha+8 \tilde{C}^{\rm{ac}}) k'^2 \right] + (1+\alpha )
4 \tilde{C}^{\rm{ac}} k'^4 =0.
\ee

This equation describes two different modes, one is a gapped sound mode, also called the inertial
mode, while the other is the acoustic Tkachenko mode of the triangular lattice. To the lowest order
in the long-wavelength approximation we get
\bea\label{modestriac}
\omega^{\rm{ac}}_I &=& 2 \Omega + \frac{ 1+ \alpha + 8 \tilde{C}^{\rm{ac}}}{4 \Omega} k'^2, \\
\nonumber \omega^{\rm{ac}}_T &=& \sqrt{(1+\alpha) \tilde{C}^{\rm{ac}}} \frac{ k'^2}{\Omega}.
\eea
The inertial mode is gapped, starting at $2 \Omega$, and the second mode is the acoustic Tkachenko
mode which has quadratic dispersion at long-wavelengths, similar to the Tkachenko mode in a
one-component vortex lattice.

Calculation of the optical Tkachenko mode, similarly, starts by evaluating the optical part of the
stress as
\be
\vec{\sigma}_- = 4 C^{\rm{op}} \vec{\epsilon}_-.
\ee
From the polarization equation we get
\be
\epsilon^y_- = \frac{N_1}{D_1} \epsilon^x_-,
\ee
with
\bea
N_1 &=&  \omega k_y - i \left(2 \Omega  + \frac{ 2 C^{\rm{op}}}{m n \Omega} \right) k_x \\
\nonumber D_1 &=&  \omega k_x + i \left(2 \Omega  + \frac{ 2 C^{\rm{op}}}{m n \Omega} \right) k_y
\eea
which results in two coupled equations obtained from Eqs. (\ref{sound}) and (\ref{lattice-sound})
\be
\left( \omega^2 - (1-\alpha) \frac{g n}{m} k^2 \right) n_- + n \left(4 \Omega^2 + \frac{4
C^{\rm{op}}}{m n} \right) i \frac{\omega k^2}{D_1} \epsilon_-^x =0,
\ee
and
\be
\frac{ \omega^2 }{n} n_- + \left[ \omega^2 - \frac{2 C^{\rm{op}}}{\Omega n m} \left( 2 \Omega +
\frac{2 C^{\rm{op}}}{\Omega n m} \right) \right] i \frac{\omega k^2}{D_1} \epsilon_-^x =0.
\ee

Once again, using $k'=\sqrt{\frac{g n}{m}} k$, and the dimensionless elastic constants we obtain
the dispersion relation for two modes,
\bea\label{modestriop}
\omega^{\rm{op}}_I &=& 2 \Omega \sqrt{ 1+ \frac{\sqrt{3}}{\pi} \tilde{C}^{\rm{op}} \frac{g n}{\Omega}}+ \frac{ 1 - \alpha }{4 \Omega} k'^2, \\
\nonumber \omega^{\rm{op}}_T &=& \sqrt{\frac{\sqrt{3}}{2\pi} \tilde{C}^{\rm{op}} \frac{g
n}{\Omega}} \: k'.
\eea
These results are obtained to the lowest nonvanishing order in $k'$ and also to the lowest order in
$\frac{g n }{\Omega}$, which is a small parameter in the LLL regime.

\begin{figure}
 \centering\includegraphics[scale=0.4]{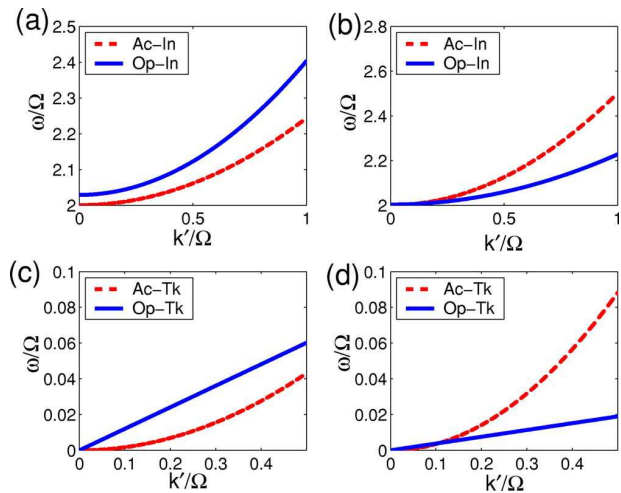}\caption[]{Spectrum for overlapped
triangular lattice, (a), (c), at $\alpha=-0.5$, and interlaced triangular lattice (b), (d) at
$\alpha=0.1$. $k'$ and $\omega$ are scaled to rotation frequency $\Omega$, and
$\frac{gn}{\Omega}=0.1$. Dispersion relations are the same for both lattice types, Eqs.
(\ref{modestriac}) and (\ref{modestriop}). However, the elastic constants are different (see Fig.
\ref{elastic1}). Both acoustic and optical inertial modes, (a), (b), are gapped. For both lattices
optical Tkachenko modes are linear while acoustic Tkachenko modes are quadratic in
$k$.}\label{modestri}
\end{figure}

The typical spectrum of the Tkachenko modes and the gapped sound modes are displayed in Fig.
\ref{modestri}. The following properties of Tkachenko modes are revealed as a result of the above
calculation.


First, we see that doubling the number of components in the BEC results in a doubling of the modes.
Because the vortex lattice oscillations are coupled to density oscillations in a compressible
fluid, there are four branches of excitation. The two inertial modes correspond to in-phase and
out-of-phase oscillations of the densities of two-components and are gapped, starting essentially
at twice the rotation frequency. As a second point, we find that the acoustic Tkachenko mode has
quadratic $k$ dependence at long-wavelength while the optical Tkachenko mode goes linearly with
$k$. In an incompressible fluid, we would expect to find the acoustic modes with linear dispersion
and the optical modes to be gapped. This result can be obtained by explicitly decoupling the
density in the above calculation. Thus the extra factor of $k$ in the dispersion is a result of the
coupling between the density and the vortex lattice oscillations.

While these two properties apply to all the lattice types considered below, there are some
properties that are specific to the overlapped triangular lattice discussed in this section. First
of all, the dispersion relation for both the optical and the acoustic Tkachenko modes are
isotropic, i.e., independent of the direction of $\vec{k}$. The isotropy of the excitations is a
direct consequence of the sixfold symmetry of the underlying lattice. The elastic (sound) waves in
a triangular lattice also show isotropic behavior \cite{lli70} and as we view the Tkachenko modes
as the elastic excitations of the vortex lattice, this result is not unexpected. However, for the
other, nontriangular, lattice types considered below, Tkachenko mode spectrum is anisotropic. A
second property is revealed by investigating the behavior of the modes for changing $\alpha$. As
$\alpha$ goes to zero, the optical Tkachenko mode becomes softer and softer, revealing that the two
lattices become mostly independent. Indeed at $\alpha=0$ there is a first-order phase transition to
the interlaced triangular lattice. As $\alpha$ approaches $-1$, this time it is the acoustic mode
that becomes soft, and there is an instability towards collapse at exactly $\alpha=-1$. It is
interesting to note that although our approach cannot describe this collapse, its signature is
still present in the Tkachenko mode spectrum.

\section{Interlaced triangular lattice}
\label{inttriangle}

For a single-component vortex lattice the equilibrium configuration is always the triangular
lattice. When the interaction between the components of a two-component vortex lattice is weak,
both vortex lattices stay triangular. The offset between the two-components is, however, decided by
the sign of the intercomponent interaction. For attractive interaction $\alpha<0$, the resulting
lattice is the overlapped triangular lattice discussed in the previous section. For weak and
repulsive interaction, it is energetically favorable to place the vortices of one-component at the
density maxima of the other component. The resulting, interlaced triangular lattice is described by
\be
u_*=\frac{1}{2}, \qquad v_*=\frac{\sqrt{3}}{2}, \qquad a_*=b_*=\frac{1}{3}.
\ee
The interlaced triangular lattice is the minimum-energy configuration for $0<\alpha<0.1724$, and is
displayed in Fig. \ref{latticegeo}.

The elastic energy and the Tkachenko mode equations follow directly from the symmetry of the
lattice. As the interlaced triangular lattice has exactly the same symmetry as the overlapped
triangular lattice discussed in the previous section, the calculation given in the previous section
is valid also for the interlaced triangular lattice. It is only the values of the elastic constants
$\tilde{C}^{\rm{ac}}$ and $\tilde{C}^{\rm{op}}$, and their dependence on $\alpha$, that is
different from the previous case.

As a result, the acoustic modes are given by
\bea
\omega^{\rm{ac}}_I &=& 2 \Omega + \frac{ 1+ \alpha + 8 \tilde{C}^{\rm{ac}}}{4 \Omega} k'^2, \\
\nonumber \omega^{\rm{ac}}_T &=& \sqrt{(1+\alpha) \tilde{C}^{\rm{ac}}} \frac{ k'^2}{\Omega}
\eea
and the optical modes are given by
\bea
\omega^{\rm{op}}_I &=& 2 \Omega \sqrt{ 1+ \frac{\sqrt{3}}{\pi} \tilde{C}^{\rm{op}} \frac{g
n}{\Omega}}+ \frac{ 1 - \alpha }{4 \Omega} k'^2, \\ \nonumber \omega^{\rm{op}}_T &=&
\sqrt{\frac{\sqrt{3}}{2\pi} \tilde{C}^{\rm{op}} \frac{g n}{\Omega}} \: k'.
\eea
Here the relations between $C^{\rm{op}},C^{\rm{ac}}$ and $C_u,C_{ab}$, remain the same as in the
previous section. A plot of the elastic constants is given in Fig. \ref{elastic1}.

As in the previous section, we see that the acoustic Tkachenko mode is quadratic in $k$, at
long-wavelengths, while the optical Tkachenko mode is linear in $k$. As a consequence of the
sixfold symmetry of the underlying lattice, both modes are isotropic. A typical spectrum of the
Tkachenko modes is displayed in Fig. \ref{modestri}. Just as the optical Tkachenko mode becomes
soft for the overlapped triangular lattice as $\alpha=0$ is approached from below, a similar
softening takes place for the interlaced triangular lattice. So both sides of the first-order
transition have dynamics characterized by a soft optical mode.

\section{Square lattice}
\label{square}

The lattice type which is energetically favorable over the largest range of intercomponent
interaction is the square lattice. For $0.3733 < \alpha < 0.9256 $, one-component's vortex lattice
forms a square lattice while the other components vortices are situated at the centers of the
squares (see Fig. \ref{latticegeo}). This lattice is characterized by
\be
u_*=0, \qquad v_*=1, \qquad a_*=b_*=\frac{1}{2}.
\ee

For the square lattice we can write the elastic energy due to optical and acoustic deformations as
\be
E_{\rm{elastic}}= E_{\rm{elastic}}^{\rm{ac}} + E_{\rm{elastic}}^{\rm{op}}
\ee
with
\bea
E_{\rm{elastic}}^{\rm{ac}} &=& \frac{1}{2} \int d^2r \left[ C_1^{\rm{ac}} \frac{ \partial
\epsilon_+^x}{\partial x} \frac{\partial \epsilon_+^y}{\partial y} + C_2^{\rm{ac}} \left(
\frac{\partial \epsilon_+^x}{\partial y} + \frac{\partial \epsilon_+^y}{\partial x} \right)^2
\right], \\ \nonumber E_{\rm{elastic}}^{\rm{op}} &=& \int d^2 r C^{\rm{op}} \left( \vec{\epsilon}_-
\right)^2.
\eea

\begin{figure}
 \centering\includegraphics[scale=0.35]{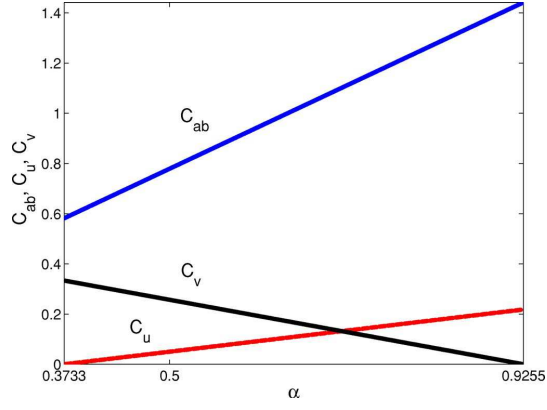}\caption[]{Elastic constants
($C_{ab}, C_u, C_v$) of square lattice, Eqs. (\ref{elassquac}) and (\ref{elassquop}). As the
components attract each other more, $C_{ab}$ increases linearly. Both limits of $\alpha$ lead to
second-order phase transitions. $C_u$ and $C_v$ vanish at $\alpha=0.3733$ and $\alpha=0.9255$,
respectively.}\label{elastic2}
\end{figure}

For acoustic modes we define the dimensionless elastic constants
\be
\tilde{C}_1^{\rm{ac}} = \frac{C_1^{\rm{ac}}}{g n^2}, \qquad \tilde{C}_2^{\rm{ac}} =
\frac{C_2^{\rm{ac}}}{g n^2},
\ee
and fit the acoustic part of the elastic energy to the form
\be
\label{elassquac}
E_{\rm{elastic}}^{\rm{ac}} = \frac{ g n^2}{2} \left[ C_u (u-u_*)^2 + C_v (v-v_*)^2 \right],
\ee
which yield
\be
\tilde{C}_{1}^{\rm{ac}}= -4 C_v, \qquad \tilde{C}_{2}^{\rm{ac}}= C_u.
\ee
The variation of elastic constants $C_u$ and $C_v$ are plotted in Fig. \ref{elastic2}.


The calculation of acoustic Tkachenko mode frequencies proceed similar to the previous sections.
However, for the square lattice, the equations are not isotropic, for example the polarization
equation (\ref{polarization}) gives
\be \epsilon_+^y=
\frac{N_1}{D_1} \epsilon_+^x,
\ee
with
\bea
N_1 &=& \left( -i \omega k_y - 2 \Omega k_x - \frac{4 C_2^{\rm{ac}} + C_1^{\rm{ac}}}{2 n m \Omega}
k_x k_y^2 \right), \\ \nonumber D_1 &=& \left( -i \omega k_x + 2 \Omega k_y + \frac{4 C_2^{\rm{ac}}
+ C_1^{\rm{ac}}}{2 n m \Omega} k_y k_x^2 \right).
\eea
We find, in the long-wavelength limit, the inertial mode frequency
\be
\omega_I^{\rm{ac}} = 2 \Omega + \frac{1+\alpha + 2 \tilde{C}_{2}^{\rm{ac}}}{4 \Omega} {k'}^2,
\ee
and the acoustic Tkachenko mode frequency
\be
\label{squac} \omega_T^{\rm{ac}} = \sqrt{\frac{1+\alpha}{2}} \sqrt{  \left[ \tilde{C}_2^{\rm{ac}}
f_2(\theta) - \tilde{C}_1^{\rm{ac}} f_1(\theta) \right])} \frac{k'^2}{ \Omega}.
\ee
Here, we have
\be
f_1(\theta)= \frac{1}{4} \sin^2(2 \theta), \qquad f_2(\theta)= \cos^2(2 \theta),
\ee
where $\theta$ is the angle from the $\hat{x}$ direction when the basis vectors of the vortex
lattice are taken along $\hat{x}$ and $\hat{y}$.

For the optical spectrum, we  define the dimensionless elastic constant
\be
\tilde{C}^{\rm{op}} = \frac{d^2}{g n^2} C^{\rm{op}},
\ee
where the lattice constant $d$ is given by $d^2=\frac{\pi}{\Omega m}$. The optical part of the
elastic energy can be numerically fitted to a form
\be
\label{elassquop}
E_{\rm{elastic}}^{\rm{op}} = \frac{ g n^2}{2}  C_{ab} \left[(a-a_*)^2 + (b-b_*)^2 \right],
\ee
which yields
\be
\tilde{C}^{\rm{op}}= \frac{C_{ab}}{2}.
\ee
The dependence of the elastic constant $C_{ab}$ on $\alpha$ is plotted in Fig. \ref{elastic2}.

The gapped inertial mode and the optical Tkachenko mode are calculated to the lowest order in $k'$
and $\frac{g n}{\Omega}$ as
\be
(\omega_I^{\rm{op}})^2 = 4 \Omega^2 \left[ 1+ \frac{2 g n}{ \pi \Omega} \tilde{C}^{\rm{op}} \right]
+ (1-\alpha) {k'}^2,
\ee
and
\be
\omega_T^{\rm{op}} = \sqrt{ \frac{1-\alpha}{\pi+ 2 \frac{g n }{\Omega} \tilde{C}^{\rm{op}}}} \sqrt{
\frac{g n}{\Omega} \tilde{C}^{\rm{op}}} \: k',
\ee
respectively.


\begin{figure}
 \centering\includegraphics[scale=0.35]{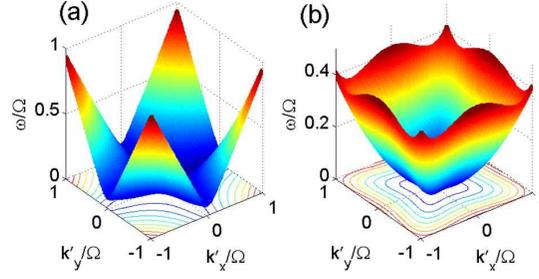}\caption[]{ Dispersion
relation of the acoustic Tkachenko modes for the square lattice, Eq. (\ref{squac}) for $\alpha=0.4$
(a), and $\alpha=0.85$ (b). Underlying contour plots are given to illustrate the anisotropy of the
modes.}\label{modesqu}
\end{figure}
The above results reveal a number of properties of the Tkachenko modes of a square vortex lattice.
Similar to the triangular lattice, there are two gapped modes, which are the sound modes of the
two-component condensate modified by the interactions with the vortex oscillations. The remaining
two gapless modes are the acoustic Tkachenko mode and the optical Tkachenko mode, which have $k^2$
and $k$ dispersion, respectively. Both the two gapped modes and the optical Tkachenko mode have
isotropic behavior, however, the underlying square lattice causes the acoustic Tkachenko mode
dispersion to be anisotropic. The anisotropy of the acoustic Tkachenko mode is more transparent
when written as
\be
\omega_T^{\rm{ac}} = \sqrt{ 2(1+\alpha) \left[ C_u \cos^2(2 \theta) + C_v \sin^2(2 \theta) \right]
} \frac{{k'}^2}{2 \Omega}.
\ee
The acoustic mode spectrum is plotted in Fig. \ref{modesqu} for two different values of $\alpha$.
We notice that depending on the elastic constants two different types of softening happens for the
acoustic Tkachenko modes. If $C_u =0 $, then the acoustic modes along the directions
\be
\theta = 0, \pi/2, \pi, 3 \pi/2
\ee
become soft. We see from the numerical fit that $C_u$ becomes zero near $\alpha=0.3733$ and these
soft modes control the dynamics of the second-order phase transition to the rhombic lattice. The
other possibility for soft mode formation is when $C_v=0$. In this case the soft acoustic modes are
along the directions
\be
\theta = \pi/4, 3 \pi/4, 5 \pi/4, 7 \pi/4.
\ee
These soft modes, then, signal the second-order phase transition to the rectangular lattice, at
$\alpha=0.9256$.

\section{Rectangular lattice}
\label{rectangle}

When the interactions between the components are close to the interactions within each component,
the energetically favorable lattice becomes a rectangular lattice. The rectangular lattice has
$a_*=b_*=1/2$, so vortices of one-component are placed at the centers of the rectangles formed by
the vortices of the other component. The ratio of the long side of the rectangles to their short
side, $v_*$, increases with increasing $\alpha$ (see  Fig. \ref{latticegeo2}).

The elastic energy can once again be separated as
\be
E_{\rm{elastic}}= E_{\rm{elastic}}^{\rm{ac}} + E_{\rm{elastic}}^{\rm{op}}.
\ee
Here, we express the acoustic part of the elastic energy as
\be
E_{\rm{elastic}}^{\rm{ac}} = \frac{1}{2} \int d^2r \left[ C_1^{\rm{ac}} \left(\frac{ \partial
\epsilon_+^x}{\partial x} - \frac{\partial \epsilon_+^y}{\partial y} \right)^2 + C_2^{\rm{ac}}
\left( \frac{\partial \epsilon_+^x}{\partial y} + \frac{\partial \epsilon_+^y}{\partial x}
\right)^2 \right],
\ee
a form that is essentially the same as the square lattice, as the hydrostatic compression modulus
is zero. The optical part is
\be
E_{\rm{elastic}}^{\rm{op}} = \int d^2 r \left[ C_1^{\rm{op}} \left( \epsilon_-^x \right)^2 +
C_2^{\rm{op}} \left( \epsilon_-^y \right)^2 \right].
\ee

\begin{figure}
 \centering\includegraphics[scale=0.35]{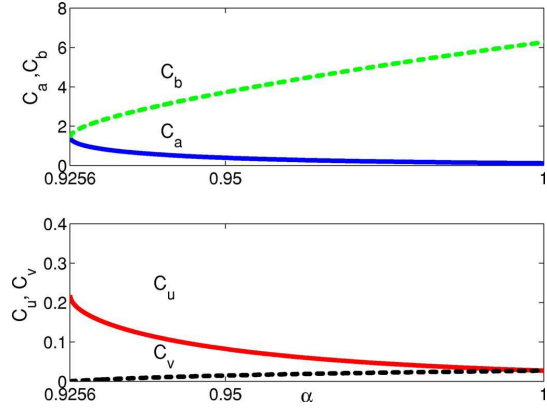}\caption[]{ Elastic constants ($C_{a},
C_b, C_u, C_v$) of rectangular lattice. The upper figure shows optical elastic constants($C_{a},
C_b$). As $\alpha \to 1$, $C_a$ vanishes. The lower figure shows acoustic elastic constants ($C_u,
C_v$). As $\alpha \to 1$, $ C_u \to C_v$ and there remains only one acoustic elastic constant
similar to the one-component triangular lattice.}\label{elastic3}
\end{figure}

For the acoustic modes, we define the dimensionless elastic constants
\be
\tilde{C}_1^{\rm{ac}} = \frac{C_1^{\rm{ac}}}{g n^2}, \qquad \tilde{C}_2^{\rm{ac}} =
\frac{C_2^{\rm{ac}}}{g n^2},
\ee
and fit the acoustic part of the elastic energy to the form
\be
E_{\rm{elastic}}^{\rm{ac}} = \frac{ g n^2}{2} \left[ C_u (u-u_*)^2 + C_v (v-v_*)^2 \right],
\ee
which results in
\be
\tilde{C}_{1}^{\rm{ac}}= v_*^2 C_v, \qquad \tilde{C}_{2}^{\rm{ac}}= v_*^2 C_u.
\ee
The numerical results for elastic constants $C_u$ and $C_v$ are given in Fig. \ref{elastic3}.


As a result, we calculate the acoustic Tkachenko mode frequency
\be
\label{recac} \omega_T^{\rm{ac}} = \sqrt{\frac{1+\alpha}{2}} \sqrt{ \tilde{C}_1^{\rm{ac}} \sin^2(2 \theta) + \tilde{C}_2^{\rm{ac}} \cos^2(2 \theta) } \frac{{k'}^2}{\Omega},
\ee
and the acoustic inertial mode frequency
\be
(\omega_I^{\rm{ac}})^2 = 4 \Omega^2 + [ 1+ \alpha + 2 (\tilde{C}_1^{\rm{ac}}+\tilde{C}_2^{\rm{ac}})
] {k'}^2.
\ee

For the optical modes, we nondimensionalize
\be
\tilde{C}_1^{\rm{op}} = \frac{d_1^2}{g n^2} C_1^{\rm{op}}, \qquad \tilde{C}_2^{\rm{op}} =
\frac{d_2^2}{g n^2} C_2^{\rm{op}},
\ee
where $d_1$ and $d_2$ are the sides of the rectangular unit cell with
\be
d_1^2 = \frac{\pi}{\Omega m v_*}, \qquad d_2^2 = \frac{ \pi v_*}{\Omega m}.
\ee
When the elastic energy is fitted to the numerical form
\be
E_{\rm{elastic}}^{\rm{op}} = \frac{ g n^2}{2} \left[ C_{a} (a-a_*)^2 + C_{b} (b-b_*)^2 \right],
\ee
we obtain
\be
\tilde{C}_1^{\rm{op}} = \frac{ C_{a}}{2}, \qquad \tilde{C}_2^{\rm{op}} = \frac{ C_{b}}{2},
\ee
The dependence of $C_a$ and $C_b$  on $\alpha$ is plotted in Fig. \ref{elastic3}.

As a result of the calculation, we obtain the optical Tkachenko mode dispersion
\be
\label{RectangularOpticalMode}
\omega_T^{\rm{op}} = \sqrt{ \frac{1-\alpha}{\pi} \frac{g n}{\Omega}} \sqrt{
\frac{\tilde{C}_2^{\rm{op}}}{v_*} \cos^2(\theta) + \tilde{C}_1^{\rm{op}} v_* \sin^2(\theta)} \: k',
\ee
and the inertial mode frequencies
\be
(\omega_I^{\rm{op}})^2 = 4 \Omega^2 \left[ 1+ \frac{ g n}{ \pi \Omega} \left( v_*
\tilde{C}_1^{\rm{op}} + \frac{1}{v_*} \tilde{C}_2^{\rm{op}} \right) \right] + (1-\alpha) {k'}^2.
\ee

\begin{figure}
 \centering\includegraphics[scale=0.25]{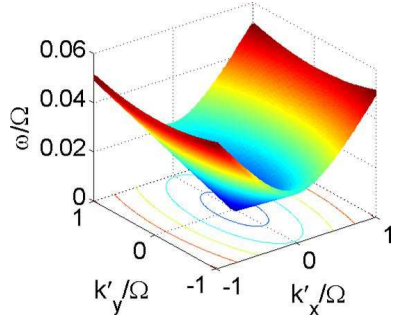}\caption[]{
Dispersion relation of the optical Tkachenko mode of the rectangular lattice, Eq.
(\ref{RectangularOpticalMode}), for $\alpha=0.95$, $\frac{gn}{\Omega}=0.1$. The underlying contour
plot reflects the symmetry of the rectangular lattice.}\label{modesrecop}
\end{figure}

A number of important conclusions can be deduced from the above results. First of all, both the
acoustic and optical Tkachenko modes are anisotropic, while the inertial modes are isotropic for
the rectangular lattice. While the anisotropy of the acoustic Tkachenko mode, is similar to the
anisotropy obtained for the square lattice, the anisotropy of the optical modes can be understood
by a different mechanism. The rectangular lattice can be thought of as alternating planes of
vortices of different components. It is easier to move the vortices in these planes, rather than
perpendicular to these planes. A typical dispersion of optical Tkachenko modes is given in Fig.
\ref{modesrecop}.


As a second property, we see that near $\alpha=0.9256$ there is a soft acoustic mode, signaling a
second-order transition to the square lattice. Thus both sides of the transition from square to
rectangular lattice have a soft acoustic mode.

\begin{figure}
\centering\includegraphics[scale=0.35]{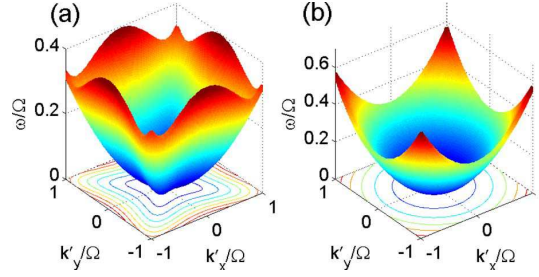}\caption[]{ Dispersion relation of the acoustic
Tkachenko modes of the rectangular lattice, Eq. (\ref{recac}) for $\alpha=0.95$ (a) and for
$\alpha=1.0$ (b). At $\alpha=1.0$, the dispersion relation becomes isotropic. The similarity
between (a) and Fig. \ref{modesqu}(b) is due to the second-order phase transition between square
and rectangular lattices.}\label{modesrecac}
\end{figure}

When intercomponent interaction is equal in strength to the interaction within the components,
i.e., $\alpha=1$, a number of interesting phenomena are expected. First, at $\alpha=1$, there is no
distinction between different components, and one would expect the results to be the same as that
of a single-component vortex lattice. Indeed, at this point, $v_*=\sqrt{3}$, and the resulting
rectangular lattice is equivalent to a single-component triangular lattice. Furthermore, the
acoustic mode spectrum becomes isotropic exactly at this point, as can be seen in Fig.
\ref{modesrecac}.


However, the point $\alpha=1$, where the intercomponent interaction is the same as the interaction
between the components, is special in another way. For a nonrotating two-component BEC, there is an
instability towards phase-separation at this point. In previous studies of two-component BECs with
vortex lattices, this instability was not observed. However, we find that at $\alpha=1$ the optical
Tkachenko mode becomes soft, and the system is unstable beyond $\alpha=1$. This is reflected in the
$\sqrt{1-\alpha}$ term, in the dispersion relation Eq. (\ref{RectangularOpticalMode}). Thus we find
that there is an instability beyond $\alpha=1$, for rapidly rotating two-component condensates. As
previous studies of this system did not take the coupling between the vortex movement and density
oscillations into account, it is not surprising that this instability was not observed.

Although we find that there is an instability at $\alpha=1$, it is not clear that this  instability
leads directly to phase-separation. The analog of the sound mode that is unstable in a nonrotating
system is the optical inertial mode. As this mode has a gap, there is no instability in the
long-wavelength. We find that the dispersion of the optical inertial mode has a $k^2$ term with a
negative coefficient, but this is not sufficient to claim that there will be an instability at a
finite value of $k$, as higher-order terms such as $k^4$ may prevent the dispersion from reaching
zero frequency. Instead there may be a phase with partial phase-separation and disordered
distribution of vortices beyond $\alpha=1$. Further investigation of this instability is needed to
determine the nature of the phase beyond $\alpha=1$.

\section{Rhombic lattice}
\label{rhombus}

The final lattice type we consider is the rhombic lattice, which is the minimum-energy
configuration for $0.1724 < \alpha < 0.3733 $. This lattice is an intervening phase between the
interlaced triangular lattice and the square lattice discussed in previous sections. At
$\alpha=0.1724$ there is a first-order transition from the interlaced triangular lattice, where
$a_*$ and $b_*$ change discontinuously from $1/2$ to
\be
a_*=b_*=\frac{1}{3}.
\ee
The unit cell also becomes a rhombus, while the acute angle of the rhombus $\eta$ continuously
changes from $67.96^\circ$ to $90^\circ$. A plot of
 the lattice geometry and the change of $\eta$ is given in Fig. \ref{latticegeo2}.

The rhombic lattice has twofold (reflection) symmetry  along the  axis that makes an angle $\eta/2$
with the primitive basis vectors. However, instead of expressly taking advantage  of this symmetry,
we use a general form for the elastic energy. Writing
\be
E_{\rm{elastic}}= E_{\rm{elastic}}^{\rm{ac}} + E_{\rm{elastic}}^{\rm{op}},
\ee
we use
\be
E_{\rm{elastic}}^{\rm{ac}} = \frac{1}{2} \int d^2r \left[ C_1^{\rm{ac}} \left(\frac{ \partial
\epsilon_+^x}{\partial x} - \frac{\partial \epsilon_+^y}{\partial y} \right)^2 + C_2^{\rm{ac}}
\left( \frac{\partial \epsilon_+^x}{\partial y} + \frac{\partial \epsilon_+^y}{\partial x}
\right)^2 + C_3^{\rm{ac}} \left( \frac{\partial \epsilon_+^x}{\partial y} + \frac{\partial
\epsilon_+^y}{\partial x} \right)\left(\frac{ \partial \epsilon_+^x}{\partial x} - \frac{\partial
\epsilon_+^y}{\partial y} \right) \right],
\ee
and
\be
E_{\rm{elastic}}^{\rm{op}} = \int d^2 r \left[ C_1^{\rm{op}} \left( \epsilon_-^x \right)^2 +
C_2^{\rm{op}} \left( \epsilon_-^y \right)^2 + C_3^{\rm{op}} \epsilon_-^x \epsilon_-^y \right].
\ee

\begin{figure}
 \centering\includegraphics[scale=0.38]{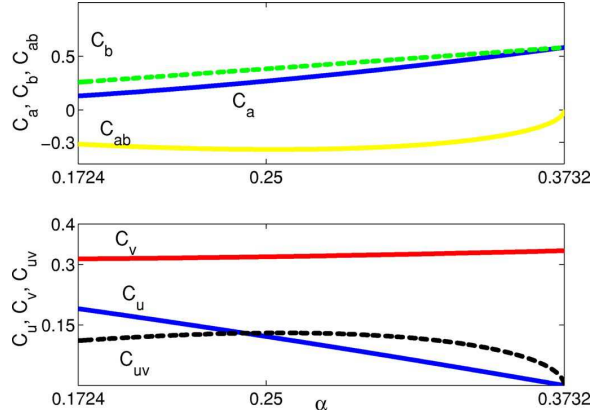}\caption[]{ Optical
elastic constants [upper, Eq. (\ref{elasrhoop})] and acoustic elastic constants [lower, Eq.
(\ref{elasrhoac})] of rhombic lattice with respect to $\alpha$. As $\alpha \to 0.3732$, $C_{a} \to
C_b$, and $C_u ,C_v $ vanish, leaving  two optical elastic constants, and  one acoustic elastic
constant for the square lattice. In the opposite limit $\alpha \to 0.1724$, six elastic constants
remain due to the discontinuity in the transition to interlaced triangular
lattice.}\label{elastic4}
\end{figure}

For the acoustic modes, we define dimensionless quantities
\be
\tilde{C}_1^{\rm{ac}} = \frac{C_1^{\rm{ac}}}{g n^2}, \qquad \tilde{C}_2^{\rm{ac}} =
\frac{C_2^{\rm{ac}}}{g n^2},
 \qquad \tilde{C}_3^{\rm{ac}} = \frac{C_3^{\rm{ac}}}{g n^2},
\ee
and fit the acoustic part of the elastic energy to the form
\be
\label{elasrhoac}
E_{\rm{elastic}}^{\rm{ac}} = \frac{ g n^2}{2} \left[ C_u (u-u_*)^2 + C_v (v-v_*)^2 + C_{uv} (u-u_*)
(v-v_*) \right],
\ee
which yields
\be
\tilde{C}_{1}^{\rm{ac}}= v_*^2 C_v, \qquad \tilde{C}_{2}^{\rm{ac}}= v_*^2 C_u, \qquad
\tilde{C}_{3}^{\rm{ac}}= - v_*^2 C_{uv}.
\ee
The numerical results for elastic constants $C_u$, $C_v$, and $C_{uv}$ are given in Fig.
\ref{elastic4}.


We find the acoustic inertial mode dispersion
\be
(\omega_I^{\rm{ac}})^2 = 4 \Omega^2 + [ 1+ \alpha + 2 (\tilde{C}_1^{\rm{ac}}+\tilde{C}_2^{\rm{ac}})
] {k'}^2,
\ee
and the acoustic Tkachenko mode dispersion
\be\label{rhoac}
\omega_T^{\rm{ac}} = \sqrt{\frac{1+\alpha}{2}} \sqrt{ 4 \tilde{C}_1^{\rm{ac}} \frac{k_x^2
k_y^2}{k^4}  + \tilde{C}_2^{\rm{ac}} \frac{(k_x^2-k_y^2)^2}{k^4} + 2 \tilde{C}_3^{\rm{ac}}
\frac{k_x k_y (k_y^2-k_x^2)}{k^4}} \frac{{k'}^2}{\Omega}.
\ee
The anisotropy of the Tkacheko mode is more transparent when  represented  in terms of $\theta$,
the angle from the $\hat{x}$ axis,
\be
\omega_T^{\rm{ac}} = \sqrt{\frac{1+\alpha}{2}} \sqrt{ \tilde{C}_1^{\rm{ac}} \sin^2(2 \theta) +
\tilde{C}_2^{\rm{ac}} \cos^2(2 \theta) - \tilde{C}_3^{\rm{ac}} \sin(4 \theta) }
\frac{{k'}^2}{\Omega}.
\ee


\begin{figure}
 \centering\includegraphics[scale=0.25]{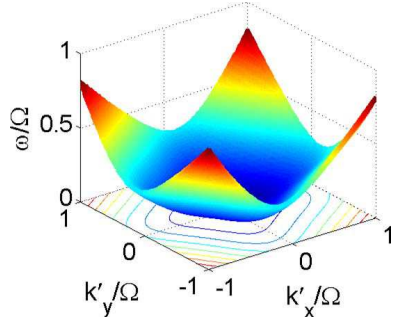}\caption[]{ Dispersion
relation of the acoustic Tkachenko mode of the rhombic lattice, Eq. (\ref{rhoac}), for
$\alpha=0.2$. The anisotropy reflects the twofold symmetry of the rhombic lattice (see Fig.
\ref{polar}).}\label{modesrho}
\end{figure}

For the optical modes, we define  the dimensionless elastic constants,
\be
\tilde{C}_1^{\rm{op}} = \frac{d^2}{g n^2} C_1^{\rm{op}}, \qquad \tilde{C}_2^{\rm{op}} =
\frac{d^2}{g n^2} C_2^{\rm{op}}, \qquad \tilde{C}_3^{\rm{op}} = \frac{d^2}{g n^2} C_3^{\rm{op}},
\ee
where the side length of the  rhombus $d$ is
\be
d^2 = \frac{\pi}{\Omega m \sin(\eta)}.
\ee
We use a numerical fit to the energy of the general form
\be
\label{elasrhoop} E_{\rm{elastic}}^{\rm{op}} = \frac{ g n^2}{2} \left\{ C_{ax} [(a-a_*)+
u_*(b-b_*)]^2 + C_{by} v_*^2 (b-b_*)^2 + C_{ab} [(a-a_*)+ u_* (b-b_*)] v_* (b-b_*) \right\},
\ee
which yields
\be
\tilde{C}_1^{\rm{op}} = \frac{ C_{ax}}{2}, \qquad \tilde{C}_2^{\rm{op}} = \frac{ C_{by}}{2}, \qquad
\tilde{C}_3^{\rm{op}} = \frac{ C_{ab}}{2},
\ee
The dependence of $C_{ax}$, $C_{by}$, and $C_{ab}$  on $\alpha$ is plotted in Fig. \ref{elastic4}.

We find the optical inertial mode
\be
(\omega_I^{\rm{op}})^2 = 4 \Omega^2 \left[ 1+ \frac{ g n}{ \pi \Omega} \sin(\eta) \left(
\tilde{C}_1^{\rm{op}} +  \tilde{C}_2^{\rm{op}} \right) \right] + (1-\alpha) {k'}^2,
\ee
and the optical Tkachenko mode
\be
\label{RhombicOpticalMode}
\omega_T^{\rm{op}} = \sqrt{ \frac{1-\alpha}{\pi} \frac{g n}{\Omega} \sin(\eta)} \sqrt{
\tilde{C}_2^{\rm{op}} \cos^2(\theta) +
 \tilde{C}_1^{\rm{op}} \sin^2(\theta) - \frac{1}{2} \tilde{C}_3^{\rm{op}} \sin(2 \theta)} \: k'.
\ee

\begin{figure}
\centering\includegraphics[scale=0.4]{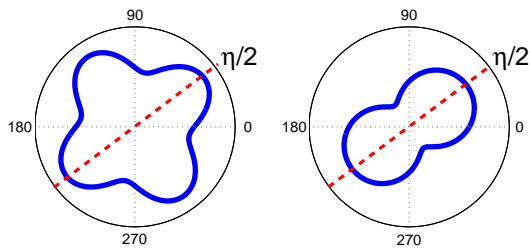}\caption[]{ Polar plot of the frequency of the
acoustic [left, Eq. (\ref{rhoac})] and the optical [right, Eq. (\ref{RhombicOpticalMode})]
Tkachenko modes for the rhombic lattice ($k=0.1$, $\frac{gn}{\Omega}=0.1$, $ \alpha=0.2$). $\eta/2$
is the opening angle of the rhombic unit cell at $\alpha=0.2$. } \label{polar}
\end{figure}

For the rhombic lattice, both the acoustic and the optical Tkachenko modes are anisotropic. A
typical dispersion for the acoustic Tkachenko modes is displayed in Fig. \ref{modesrho}. In the
calculation above we have not implicity assumed the twofold symmetry of the rhombic lattice,
however, the resulting dispersion relations respect this symmetry. As an example the polar plot of
optical and acoustic mode frequencies is given in Fig. \ref{polar}. This symmetry can be viewed as
a validation of the numerical approach we use to calculate the elastic coefficients.

Another property of the Tkachenko modes of the rhombic lattice is that the transition to the square
lattice at $\alpha=0.3733$ is accompanied by a soft acoustic mode. However, the first-order
transition to the triangular lattice does not have any soft acoustic or optical Tkachenko mode.


\section{Structural phase transitions}
\label{phasetransitions}

The  two-component vortex lattice system has five different equilibrium lattice types and four
structural phase transitions between them. In this section, we comment on  the interplay between
these transitions and the Tkachenko modes of the lattice. Our aim is to give an overall picture of
the physics in this system as the intercomponent interaction $\alpha$ is varied. We start from
$\alpha=-1$, and consider all transitions as $\alpha$ is increased.

For $\alpha<-1$, the two-component BEC system is unstable towards collapse due to the strength of
attraction between the components. This instability is apparent in the Tkachenko mode spectrum of
the overlapped triangular lattice for $\alpha$ values greater than but close to $-1$. Here there is
a soft acoustic Tkachenko mode, as discussed in Sec. \ref{ovrtriangle}.

At $\alpha=0$, when there is no interaction between the two-components of the BEC, the vortex
lattice geometry changes from overlapped triangular lattice to the interlaced triangular lattice.
This first-order transition leaves the unit cell geometry the same, however, there is a
discontinuous jump in the relative positions of vortices within the unit cell. On both sides of the
transition there is a soft optical Tkachenko mode. Thus the reordering inside the unit cell is
accompanied by a soft long-wavelength mode as expected.

As the intercomponent repulsion is increased further, there is a first-order phase transition from
the interlaced triangular lattice to the rhombic lattice, at $\alpha=0.1724$. In this transition,
both the unit cell geometry and the positions of vortices inside the unit cell change
discontinuously. We find no signature of this transition in the long-wavelength  optical or
acoustic modes. The instability mechanism causing this transition must include both optical and
acoustic Tkachenko modes, and must take place at wavelengths comparable to the lattice spacing.
Thus this instability is not captured by our linear, long-wavelength approach.

Between $\alpha=0.1724$ and $\alpha=0.3733$ the rhombic lattice is the minimum-energy
configuration, and at $\alpha=0.3733$ there is a second-order phase transition to the square
lattice. On both sides of this transition there is a soft acoustic Tkachenko mode. The acoustic
modes have anisotropic dispersion for both square and rhombic lattices, and the soft mode has a
wave vector $\vec{k}$, that is parallel to the primitive lattice basis vectors $\vec{a}_1=d
\hat{x}$, or $\vec{a}_2=d \hat{y}$. As in the structural phase transitions of solids, a
second-order phase transition is accompanied by a soft acoustic mode.

The final structural phase transition between different lattice geometries takes place at
$\alpha=0.9256$, between the square and rectangular lattices. This is a second-order phase
transition, and we find that there is a soft  acoustic Tkachenko mode on both sides of the
transition. The soft mode has a wave vector $\vec{k}$ that makes an angle of $\pi/4$ with the
primitive lattice basis vectors, $\vec{k} \parallel (\vec{a}_1+\vec{a}_2)$.

For a nonrotating system there is a phase-separation instability at $\alpha=1$. We find that at
this point the optical Tkachenko modes of the rotating system become soft. However, as discussed in
Sec. \ref{rectangle}, it is not clear if this instability directly leads to phase-separation, or to
another phase.

\section{Conclusion}
\label{conclusion}

We conclude by giving a summary of our main results and  discuss their relevance to the recent
experiment on the vortex lattice of the two-component rubidium BEC.

We considered a rapidly rotating two-component BEC, and calculated the Tkachenko mode dispersion
relations for different lattice geometries. We find that a two-component vortex lattice has two
branches of Tkachenko modes, which we call acoustic and optical Tkachenko modes in analogy with
phonons. The acoustic Tkachenko modes have $k^2$ dispersion at long-wavelengths while the optical
Tkachenko modes have linear, $k$, dispersion. For all lattice types other than triangular lattices,
the dispersion relations are anisotropic. By investigating the behavior of Tkachenko modes near
structural phase transitions, we identified the soft modes that are responsible for the phase
transitions. Out of the four structural phase transitions two are of second-order, while the
remaining two are first-order. The second-order transitions are accompanied by the softening of an
acoustic mode. For one of the first-order phase transitions we identified a soft optical Tkachenko
mode, while for the other first-order transition, no such long-wavelength mode was found. We also
found that if the intercomponent repulsion is stronger than the interactions within each component,
the vortex lattice is unstable. This instability may lead to phase-separation, as is the case for a
nonrotating two-component BEC.

In a recent experiment at JILA \cite{sce04b}, a rapidly rotating two-component Rb condensate was
created. It was found that the equilibrium vortex lattice configuration is square. Furthermore,
when the lattice was perturbed, a Tkachenko like mode was observed, however, this mode was found to
be heavily damped, thus it has not been possible to measure the Tkachenko mode frequencies.

There are three important points to consider when comparing our results with this experiment. First
the interaction parameters for the Rb system used in the experiment are different from what was
considered in this paper, most importantly, interaction parameters within each component are not
the same,
\be
g_{11} \neq g_{22}.
\ee
In this case, one would expect the acoustic and optical Tkachenko modes to be coupled. However,
this coupling should be relatively small, as
\be
\frac{2(g_{11}-g_{22})}{g_{11}+g_{22}} \approx 0.05.
\ee
When the interaction strengths within each component is different, we may redefine
\be
\alpha = \frac{g_{12}}{\sqrt{g_{11} g_{22}}},
\ee
which for the Rb system is very close to 1. Although the calculations in the LLL indicate that a
rectangular lattice is more favorable, experimentally the lattice structure is found to be a square
within experimental error. This implies, as a second point, that one must take into account that
the experimental system is not fully in the LLL regime.  As the third and final point, the
experimental system is of finite extent. The overall density profile in the system is affected by
the finite size of the system and may cause in shifts in vortex positions \cite{bpe04}. More
importantly, the coupling between vortex oscillations and the density modes, coupled with other
loss mechanisms, damp the Tkachenko modes.

The above limitations prevent a direct quantitative comparison of data with the theory presented in
this paper. There are, however, some important qualitative conclusions that can be drawn. A
puzzling result of the experiment is that the Tkachenko excitations in the two-component BEC are
more heavily damped compared to a single-component system. It is thought that the main damping
mechanism is the coupling to surface modes near the edges of the cloud, but this mechanism would be
independent of whether one is using a one-component or a two-component condensate. We believe that
there are two effects that contribute to this apparently high damping rate. The method used in the
experiment to excite Tkachenko modes is to focus a resonant laser beam onto the center of the
condensate. This method excites Tkachenko modes isotropically, giving equal weight to every
direction. However, our calculations show that Tkachenko modes in a square (or rectangular) lattice
are anisotropic. This anisotropy is very pronounced if the system is close to square to rectangular
structural phase transition, which the experimental system could be as indicated by the ratio of
its interaction strengths. When Tkachenko modes are excited isotropically, because oscillations
along different directions have different frequencies, there will be a significant dephasing
effect. We believe a significant part of the observed damping is due to this dephasing. A second
effect is that, because of the coupling between the acoustic and the optical modes, during the
excitation also optical Tkachenko modes are excited. By making measurements on the vortex positions
of one-component it is not possible to distinguish one type of oscillation from the other. We
believe, if the excitation mechanism can be made anisotropic, for example by using a resonant laser
with an elliptical focus, it should be possible to observe smaller damping rates.

It is also interesting to note that it should be possible to measure optical Tkachenko modes, using
the same interference technique used in the experiment to prove that the vortices form interlaced
lattices. An optical Tkachenko mode, once excited, would cause oscillations in the visibility of
the obtained ``vortex lattice interference'' fringes.

\begin{acknowledgements}
This work was supported by TUBITAK-KARIYER Grant No. 104T165 and a TUBA-GEBIP grant. M.\"O. Oktel
would like to thank  Erich Mueller for useful discussions and Aspen Center for Physics for their
hospitality.
\end{acknowledgements}

\end{document}